\documentclass[aps,prb,twocolumn,showpacs,superscriptaddress,groupedaddress]{revtex4-1}
\usepackage{amsmath}    
\usepackage{graphicx}   
\usepackage{verbatim}   
\usepackage{color}      
\usepackage{subfigure}  

\begin{document}
\title{Experimental Proof of Universal Conductance Fluctuation in Quasi-1D Epitaxial Bi$_{2}$Se$_{3}$ Wires}

\author{Sadashige Matsuo}
\email{smatsuo@scl.kyoto-u.ac.jp}
\affiliation{Institute for Chemical Research, Kyoto University, Uji, Kyoto 611-0011, Japan}
\author{Kensaku Chida}
\affiliation{Institute for Chemical Research, Kyoto University, Uji, Kyoto 611-0011, Japan}
\author{Daichi Chiba}
\affiliation{Institute for Chemical Research, Kyoto University, Uji, Kyoto 611-0011, Japan}
\author{Keith Slevin}
\affiliation{Department of Physics, Osaka University, Toyonaka, Osaka 560-0043, Japan}
\author{Tomi Ohtsuki}
\affiliation{Department of Physics, Sophia University, Tokyo Chiyoda-ku 102-8554, Japan}
\author{Kensuke Kobayashi}
\affiliation{Department of Physics, Osaka University, Toyonaka, Osaka 560-0043, Japan}
\author{Teruo Ono}
\affiliation{Institute for Chemical Research, Kyoto University, Uji, Kyoto 611-0011, Japan}

\author{Cui-Zu Chang}
\affiliation{Beijing
National Laboratory for Condensed Matter Physics, Institute of Physics, Chinese
Academy of Sciences, Beijing 100190, China}
\author{Ke He}
\affiliation{Beijing
National Laboratory for Condensed Matter Physics, Institute of Physics, Chinese
Academy of Sciences, Beijing 100190, China}
\author{Xu-Cun Ma}
\affiliation{Beijing
National Laboratory for Condensed Matter Physics, Institute of Physics, Chinese
Academy of Sciences, Beijing 100190, China}
\author{Qi-Kun Xue}
\affiliation{Beijing
National Laboratory for Condensed Matter Physics, Institute of Physics, Chinese
Academy of Sciences, Beijing 100190, China}

\begin{abstract}
We report on conductance fluctuation in quasi-one-dimensional wires made of epitaxial Bi$_{2}$Se$_{3}$ thin film.
We found that this type of fluctuation decreases as the wire length becomes longer and that the amplitude of the fluctuation is well scaled to the coherence, thermal diffusion, and wire lengths, as predicted by conventional universal conductance fluctuation (UCF) theory.
Additionally, the amplitude of the fluctuation can be understood to be equivalent to the UCF amplitude of a system with strong spin-orbit interaction and no time-reversal symmetry.
These results indicate that the conductance fluctuation in Bi$_{2}$Se$_{3}$ wires is explainable through UCF theory.
This work is the first to verify the scaling relationship of UCF in a system with strong spin-orbit interaction.
\end{abstract}

\maketitle
\section{Introduction}
Bi$_{2}$Se$_{3}$ is a well-known and long-studied material with excellent thermoelectric properties that
consists structurally of a layered semiconductor with a narrow band gap and strong spin-orbit interaction.
Bi$_{2}$Se$_{3}$ has recently seen a resurgence in interest as a typical material of three dimensional topological insulators (3DTIs), which have spin-polarized Dirac electrons on the surface and a bulk band gap~\cite{hasanrmp2010}. Previously, 3DTIs have been theoretically predicted~\cite{Kaneprl2005,Fuprb2007} and experimentally produced~\cite{Hsiehnature2008}.
Because its surface state is protected against time-reversal-invariant perturbation, it is anticipated that 3DTIs will be used in spintronics~\cite{Yazyevprl2010} and quantum computing~\cite{Akhmerovprl2009}.
In order to facilitate such future applications and better understand the properties of surface Dirac electrons as well as electrons with strong spin-orbit interactions in the layered semiconductor, it is important to analyze the quantum transport properties of the material.

Several groups have already reported theoretical or experimental results relevant to characteristics of quantum transport phenomena, such as the weak antilocalization (WAL) effect, in intrinsic or doped Bi$_{2}$Se$_{3}$~\cite{checkelskyprl2009, checkelskyprl2011,quprl2011,chenprb2011,wangprb2011, Steinbergprb2011,Chenprl2010,Liuprb2011,Kimprb2011,onoseapex2011,taskinprl2012, baoscirep2012, Luprb2011}.
However, there is little available information on  conductance fluctuation in Bi$_{2}$Se$_{3}$~\cite{checkelskyprl2009, checkelskyprl2011, matsuoprb2012, liscirep2012}, even though universal conductance fluctuation (UCF)  \cite{stone85,lee85,altshuler85,lee87,Akkermans} is an important phenomenon in quantum transport that has attracted a good deal of theoretical interest \cite{Rossiprl2012, adroguernjp2012}.
Additionally, as the pioneering work on conductance fluctuation in Bi$_{2}$Se$_{3}$~\cite{checkelskyprl2009} reported that neither the amplitude nor the correlation function behavior of conductance fluctuation can be explained by UCF, its origin in Bi$_{2}$Se$_{3}$ remains controversial.
 In addition, there have been no quantitative reports on the amplitude scaling relationship of UCF in systems such as Bi$_2$Se$_3$ with strong spin-orbit interaction.
A quantitative study of  conductance fluctuation in Bi$_2$Se$_3$ is therefore essential to understanding the fundamental coherence physics of this material.

Previously, we reported on the temperature dependence of the amplitude of fluctuation in Bi$_2$Se$_3$ and compared the coherence length derived from the WAL effect with that from UCF~\cite{matsuoprb2012}.
While these results confirmed that UCF theory could be used to understand conductance fluctuation, several issues remain to be addressed:  the coherence length derived from the WAL effect is not perfectly consistent with that derived from UCF; and we also failed to confirm that the amplitude of the conductance fluctuation is consistent with the value predicted by conventional UCF theory.
\begin{figure}[t]
\begin{center}
\includegraphics[width=.95\linewidth]{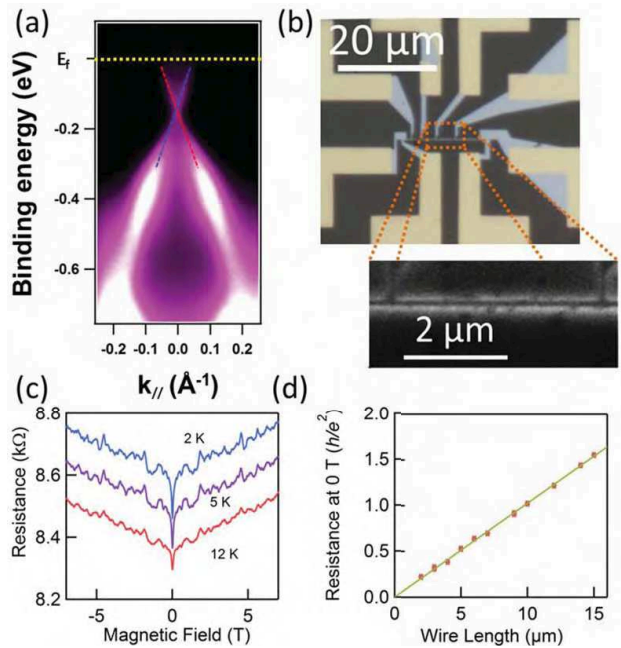}
\end{center}
\vskip-\lastskip \caption{(a) ARPES spectra of epitaxially grown 20 nm-thick film of Bi$_{2}$Se$_{3}$~\cite{matsuoprb2012}.
(b) Upper picture: Optical image of the 100 nm-width wire sample. The scale (white bar) is 20 $\rm{\mu}$m. The wire has two electrodes to inject current and six electrodes to probe voltage differences. Lower picture: SEM image showing an enlarged section of the 100 nm-width wire sample. The scale (white bar) is 2 $\rm{\mu}$m.
(c) Resistances as a function of magnetic field in the 100 nm-width and 3 $\mu$m-length wire at 2, 5, and 12 K. There are two characteristic features: the dip caused by the WAL effect; and the magnetoresistance fluctuation. The vertical axis shows the data at 2 K; the other data are incrementally shifted downward by 100 $\Omega$ for clarity.
(d) Resistance at 0 T  and 12 K as a function of wire length. The points represent experimental results and the line gives the result of  linear fitting with respect to the wire length.
} \label{fig1}
\end{figure}

In this study we report on conductance fluctuation in quasi-one-dimensional (quasi-1D) wire samples composed of an epitaxial Bi$_{2}$Se$_{3}$ thin film.
Our purpose is to clarify the physical properties of the conductance fluctuation observed in this material by addressing the scaling behavior of such fluctuations. In order to eliminate the averaging effect of wire width, in which the amplitude of fluctuation is reduced as width increases, we fabricated  quasi-1D wires, and in our results
we were clearly able to observe conductance fluctuation and the WAL effect in wires of various lengths.
Analysis of the WAL effect determined the coherence length, from which we obtained the dependence of
conductance fluctuation on the ratio of the coherence, thermal diffusion, and wire lengths,
a relationship that is  consistent with  scaling behavior expected from UCF in the case of  strong spin-orbit interaction.
This work is the first to prove the validity of conventional UCF physics based on wire, coherence, and thermal diffusion lengths in  systems with  strong spin-orbit interaction and no time-reversal symmetry; in addition,
our results confirm that UCF theory is valid when applied to Bi$_{2}$Se$_{3}$ systems.

\section{Experiment}
We grew a 20 nm-thick Bi$_2$Se$_3$ thin film by molecular beam epitaxy (MBE) on a sapphire substrate, as described previously~\cite{changspin2011}.
We used the same thin film as in the previous report~\cite{matsuoprb2012}.
The thickness of the film was equivalent to 20 quintuple layers, which is enough to allow the surface state to emerge~\cite{Zhangnatphys2010, Kimprb2011}.
Angle-resolved photoemission spectroscopy (ARPES) measurement was used to evaluate the electron band structure below the Fermi energy level in momentum-energy space, as shown in Fig.~\ref{fig1}(a)~\cite{matsuoprb2012}. 
This result shows the linear dispersion relation of the Dirac electrons on the surface, guaranteeing that our thin film is 3DTI.
The Fermi energy level is located 150 meV above the Dirac point and nearly in the bottom edge of the bulk conduction band owing to natural doping by Se vacancies~\cite{xianatphys2009}.

We deposited a 30 nm-thick amorphous Se layer {\it in situ} to protect the  Bi$_{2}$Se$_{3}$ against water or oxidation that could decrease electron mobility~\cite{butchprb2010,analytisprb2010}.
We fabricated wide and narrow wires using electron beam lithography. 
The wide one and the narrow one have 10 $\mu$m width and 100 nm width, respectively. 
Then we deposited Ti (5 nm) and Au (100 nm) as electrodes. 
Figure~\ref{fig1}(b) shows an optical image of the 100 nm-wide wire and a scanning electron microscope (SEM) image of an enlarged section of the wire; as can be seen, the edge of the wire is smooth enough to safely assume that its width is constant.
In order to determine the thin film parameters, we measured the magnetic-field dependence of the longitudinal and Hall resistances of the 10 $\mu$m width wire (Hall bar) at 13 different temperatures between 2 and 20 K.
We also measured the magnetic-field dependence of the resistance of the 100 nm-width wire at 2, 5, and 12 K.
The 100 nm-width wire had six electrodes, spaced at intervals of 1, 2, 3, 4, and 5 $\mu$m, respectively, as shown in Fig.~\ref{fig1}(b) to detect voltage difference, and two electrodes to inject current.

We obtained experimental results from wires with lengths varying from 2 to 15 $\mu$m by changing voltage probe combinations. All  measurements were carried out using the standard lock-in technique.

\section{Results and Discussions}
\subsection{Characterization of Bi$_{2}$Se$_{3}$ thin film}
Based on results obtained from the 10 $\mu$m-width Hall-bar sample, we were able to determine several relevant parameters. The carrier is n-type and the carrier density is $2.6\times 10^{13}~{\rm cm^{-2}}$, which does not depend on temperature below 20 K. 
Additionally, the temperature dependence of the resistance was found metallic.
\begin{figure}[t]
\begin{center}
\includegraphics[width=.95\linewidth]{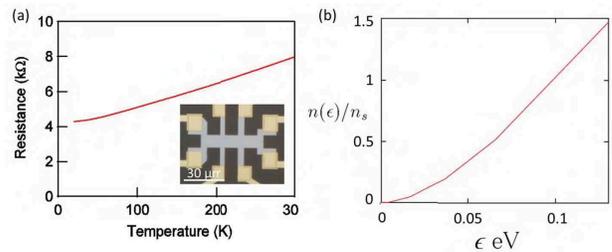}
\end{center}
\vskip-\lastskip \caption{(a)The resistance as a function of temperature between 20 K and 300 K. The resistance shows metallic behavior. The inset is the optical image of the Hall-bar sample. (b)The ratio of $n(\epsilon )$ to $n_s$. The value of $\epsilon $ corresponding to $n(\epsilon)/n_s=1$ is the Fermi energy of our thin film.
}
\label{figtempdep}
\end{figure}
We show the resistance as a function of temperature between 20 K and 300 K in Fig. \ref{figtempdep}(a).
The bulk gap of  Bi$_2$Se$_3$ is about 0.3 eV, so that, if the Fermi energy was located in the bulk gap, the temperature dependence should be insulating.
This means that our thin film was heavily n-doped semiconductor and bulk conduction mainly contributes to the transport.
Based on the temperature dependence of the carrier density and resistance, we conclude that the thin film is metallic and its carriers mainly come from the bulk conduction band.
This can be explained by the n-doping due to the oxidation in the air, the contamination on the microfabrication process or band bending near the interface by the substrate \cite{changspin2011, Zhangnatphys2010}.
For this reason, we could use a parabolic dispersion relation to calculate several relevant parameters such as the diffusion coefficient and the thermal diffusion length as follows.

\subsection{Derivation of the thermal diffusion length}
The mobility and the diffusion coefficient, $D$, are $560~{\rm cm^2}/$V$\cdot$s and $4.2\times 10^{-3}~{\rm m^2/s}$, respectively; to calculate the diffusion coefficient, we used the effective mass
0.23 $m_\mathrm{e}$~\cite{stordeurpss1992}.
The thermal diffusion length  $\sqrt{\hbar D/k_B T}$ is 127, 80, and 52 nm at 2, 5, and 12 K, respectively.
Due to finite thickness of the sample, electrons occupy several subbands, each of which has
different Fermi wave number. 
We take into account the subband occupation, and estimate $D$ from the weighted average over the subbands as seen below.

First, we calculated the subband energies. The thickness is $t=20$ nm and the subband energy can be written by 
\[
E(n) = \frac{h^2}{8m_{{\rm eff}}t^2}n^2\,,
\]
where $n$ represents the index of subband. 
The Fermi wavenumber for each subband is
\[
k_F(n, \epsilon ) = \sqrt{2\pi  (\epsilon -E(n))\rho _{2D}}\,,
\]
with $\rho _{2D} (=4\pi m_{{\rm eff}}/h^2)$  the 2D density of state per area, and $\epsilon $  the Fermi energy.

To determine how many subbands are filled for the experimentally estimated carrier density $n_s$,
we express the carrier density as a function of energy,
\begin{eqnarray}
n(\epsilon ) = \sum _{n=1}^\infty (\epsilon -E(n)) \rho _{2D}\,  \Theta(\epsilon-E(n))\,, \label{eqn_s}
\end{eqnarray}
where we have assumed zero temperature.  $\Theta(x)$ is the step function.
The Fermi energy $\epsilon$ of our thin film is obtained by equating
$n(\epsilon)=n_s=2.6\times 10^{13}$ {\rm cm$^{-2}$}.
Figure \ref{figtempdep}(b) shows the result of calculation of $n(\epsilon )/n_s$, from which
the Fermi energy of our thin film is estimated as $\epsilon=E_F=0.098$ eV,
which is smaller than the $E(5)=0.102$ eV but greater than $E(4)=0.065$ eV.
This means that subbands with $n \leq 4$ are occupied.
From the Fermi energy, we can calculate the occupation of subbands as $p(n)=(E_F-E(n)) \rho_{2D}/n_s$ and obtain $p(1)=0.348, p(2)=0.303, p(3) =0.227$ and $p(4) =0.122$.

The Fermi wavenumber used to estimate the diffusion constant is now defined by weighted average
over subbands,
\begin{eqnarray}
K_F &=& \sqrt{\sum ^4_{n=1} p(n)k_F^2(n, E_F)} \\
&=& 6.76\times 10^8  ({\rm m^{-1}}).
\end{eqnarray}
The scattering time, $\tau =7.32\times 10^{-14}$ s, is estimated from the mobility $\mu$ via
$\mu =e\tau/m_{{\rm eff}}$.
From $K_F$ and $\tau$, the diffusion constant is estimated as
\[
D=\frac{1}{2}(\frac{hK_F}{2\pi m_{{\rm eff}}})^2\times \tau = 4.2 \times 10^{-3}  ({\rm m^2/s})
\] 

We used this diffusion constant to calculate the thermal diffusion length.

\subsection{Coherence length of the wide Hall-bar sample}
In this paper, we have discussed the quantum interference effect of Bi$_2$Se$_3$, so that it is helpful to refer to the coherence length in the two-dimensional system of Bi$_2$Se$_3$.
We measured the weak anti-localization effect of the wide Hall-bar sample, whose width is 10 $\mu $m, between 2 K and 20 K. 
\begin{figure}[t]
\begin{center}
\includegraphics[width=.95\linewidth]{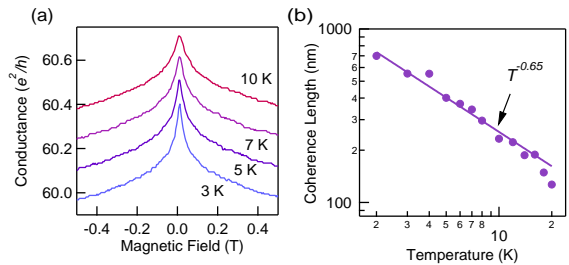}
\end{center}
\vskip-\lastskip \caption{(a) The magnetoconductance at 3, 5, 7 and 10 K. The vertical axis shows the data at 3 K; the other data are incrementally shifted upward by 0.1$e^2/h$ for clarity.
(b)  Temperature dependence of the coherence length. The solid line shows the fit to the power function. The exponent of the power function is -0.65.
}
\label{figchr}
\end{figure}
In Fig. \ref{figchr}(a), the magnetoconductance around 0 T at 3, 5, 7 and 10 K are shown. The vertical axis shows the data at 3 K; the other data are incrementally shifted upward by 0.1$e^2/h$ for clarity.
Then, we deduced the coherence length, $l_{\phi }$, using Hikami-Larkin-Nagaoka formula (HLN formula) \cite{HLN}.
\begin{equation}
 \label{eq:1}
 \begin{array}{rll}
\delta G_{\rm{WAL}}(B) &\equiv & G(B)-G(0) \\
&=& \alpha \frac{e^2}{2\pi ^2\hbar}[\psi (\frac{1}{2} + \frac{B_\phi }{B})
 -\ln (\frac{B_\phi }{B})]. \\
\end{array}
\end{equation}
$B_\phi $ is defined by $\hbar/4el_\phi ^2$.
Figure \ref{figchr}(b) represents the coherence length as a function of temperature. The dots show the coherence length deduced from the WAL effect. The solid line is the fit to a power function, whose exponent is -0.65. 
In two dimensional system, the coherence length as a function of temperature is theoretically explained by $T^{-0.5}$ if the decoherence mechanism is dominated by electron-electron interaction \cite{altshulerjpc1982}.
Our result is quite similar to the predicted value.

Our estimate of  the prefactor in Eq. (\ref{eq:1}) is $\alpha\approx -0.33$. In the case of a two-dimensional system with strong spin-orbit interaction, $\alpha$ should be equal to -0.5, which is consistent with our result.
Another group reported that this prefactor $\alpha $ becomes equal to $-1$ as the coupling between the bulk and surface states becomes smaller, because the upper surface state is not hybridized to the bulk state and contribute to the conductance independently \cite{Steinbergprb2011}.
Our estimate of $\alpha$ indicates the main contribution is from the bulk states.

\subsection{Wire-length dependence of the fluctuation amplitude}
\begin{figure}[t]
\begin{center}
\includegraphics[width=.95\linewidth]{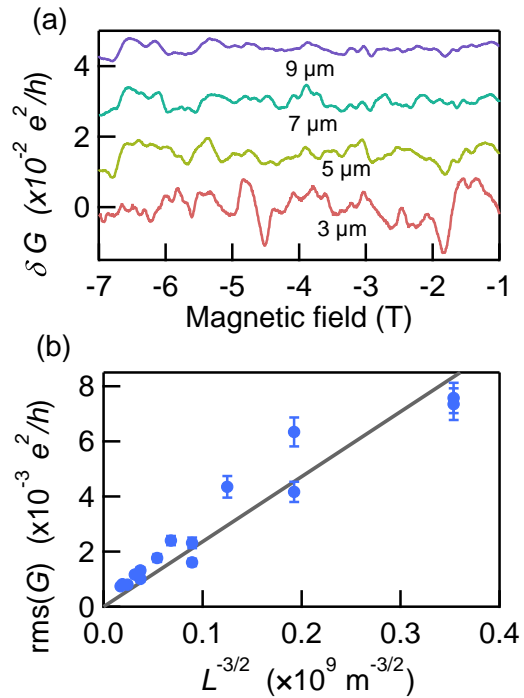}
\end{center}
\vskip-\lastskip \caption{(a) Extracted conductance fluctuation as a function of magnetic field for 100 nm-width wires of length 3, 5, 7, and 9 $\mu$m  at 2 K.
The vertical axis shows the data for the 3 $\mu$m-length wire; the other data are incrementally shifted upward by 0.015 $e^2/h$ for clarity.
(b) Amplitude of  conductance fluctuation, ${\rm rms}(G$), as a function of $L^{-3/2}$ at 2 K. The points and solid line represent the experimental data and the result of linear fitting, respectively. The value of rms($G$) is proportional to $L^{-3/2}$.
}
\label{fig2}
\end{figure}
\begin{figure}[t]
\begin{center}
\includegraphics[width=.95\linewidth]{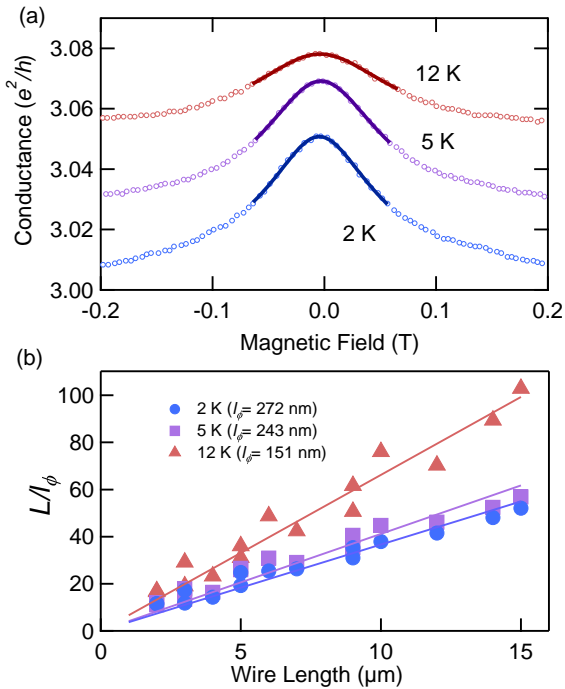}
\end{center}
\vskip-\lastskip \caption{
(a) Experimental results of conductance peak at 0 T for 100 nm-width wire at 2, 5, and 12 K are shown by the circles. The results of fitting using the model for the WAL effect in a quasi-1D system are shown as solid curves. The vertical axis shows the data at 2 K; the other data are incrementally shifted upward by 0.02 $e^2/h$ for clarity.
(b) The ratios of the wire length to the coherence length estimated from the above procedure.
These are plotted as a function of the wire length at 2, 5, and 12 K.
The linear coefficients provide the coherence lengths based on $1/l_{\phi }$.
}
\label{fig3}
\end{figure}
Results were then obtained for the 100 nm-width wire, as can be seen in Fig.~\ref{fig1}(c), which shows the magnetic field dependence of the resistance of the 3 $\mu$m-length wire at 2, 5, and 12 K.
Two important features of the resistance can be noted from the figure:
there is a dip at around 0 T originating from the WAL effect;
and there is magnetoresistance fluctuation in the entire magnetic field region.
Figure~\ref{fig1}(d) shows the relationship between the resistances at 0 T and 12 K and the wire length;
the points and the solid line represent the experimental data and the result of their linear fitting to wire length, respectively. From the figure it can be clearly seen that the samples satisfy Ohm's law on average and thus can be regarded as a system in weak disorder limit.
This linearity guarantees that they are uniform enough to investigate their wire-length dependence of the fluctuation in the resistance.

In mesoscopic samples, UCF, namely, the conductance fluctuation of the order of $e^2/h$ occurs due to the difference of how the electron wave function interferes when the electron passes through the sample
 \cite{stone85,lee85,altshuler85,lee87,Akkermans}.
The magnetic field can add the phase via the vector potential and can change the interference, causing the conductance fluctuation as a function of magnetic field.
For instance, when the phase coherence length of a quasi-1D system is finite and smaller than the sample length, self-averaging will reduce the amplitude of fluctuation by the ratio $(l_\phi /L)^{3/2}$ of the coherence length, $l_\phi $, to the wire length, $L$. When the thermal diffusion length, $l_T$, is smaller than the coherence length, $l_T < l_\phi $, the amplitude  will  be reduced through energy averaging by the ratio $l_T/l_\phi $.
In the case of a quasi-1D system with $l_T \ll l_\phi  \ll L$, the amplitude of UCF, $\Delta G=\mathrm{rms}(G)$, follows~\cite{Akkermans, beenakkerssp1991}
\begin{eqnarray}
{\rm rms}(G) = C \frac{e^2}{h}  \left(\frac{l_T {l_{\phi}}^{1/2}}{{L}^{3/2}}\right)\,,
\label{eq1}
\end{eqnarray}
where $C=\sqrt{\pi/3}$ in the presence of spin-orbit scattering and under broken time reversal symmetry.
 Below, we confirm that the conductance fluctuation in Bi$_2$Se$_3$ is proportional to $l_T l_{\phi}^{1/2}/L^{3/2}$ and estimate the proportionality coefficient $C$ in order to verify  the origin of  conductance fluctuation in Bi$_2$Se$_3$.

As can be seen in Fig.~\ref{fig1}(c), the resistance in a sample fluctuates as the magnetic field is swept;
we found that this fluctuation is reproducible regardless of the magnetic sweep direction.
In order to isolate the fluctuation component, $\delta G$, for analysis, we subtracted the linear and parabolic components of the magnetoresistance as background.
As examples, $\delta G$ of the 100 nm-width wires of lengths 3, 5, 7, and 9 $\mu$m  at 2 K is shown in Fig.~\ref{fig2}(a), in which it can be seen that the amplitude of  fluctuation becomes smaller as the wire becomes longer.
In order to ensure that this tendency is consistent with conventional UCF theory, a plot of  rms$(G)$ at 2 K as a function of $L^{-3/2}$ in Fig.~\ref{fig2}(b)
shows that the amplitude is indeed proportional to $L^{-3/2}$, as predicted by the theory.

\subsection{Coherence length of quasi-1D wires}
\begin{figure}[t]
\begin{center}
    \includegraphics[width=.5\linewidth]{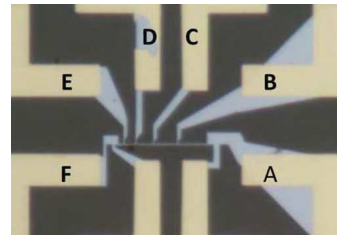}
\end{center}
  \vskip-\lastskip \caption{
   Probe configurations are shown in its optical image. The photo is the same as that shown in Fig. 1(b).
  }
  \label{figs1}
\end{figure}

To get more quantitative results about the amplitude of the fluctuation, we analyzed the conductance peak around 0 T, which comes from the WAL effect, and deduce the coherence lengths.
The circles in Fig.~\ref{fig3}(a) show the experimentally obtained conductance of the 3 $\mu$m-length wire at 2, 5, and 12 K as a function of the magnetic field near 0 T.
We then fitted the data to the WAL formula for a quasi-1D system~\cite{Akkermans}:
\begin{equation}
\label{eq2}
\delta G_{\rm{WAL}} (B)= \frac{e^2}{h}\frac{l_\phi }{L} \left(
1+\frac{4 \pi^2}{3}(wLB/\phi_0)^2\left(\frac{\l_\phi}{L}\right)^2
\right)^{-1/2}\,,
\end{equation}
where $\phi_0=h/e$ is the flux quantum and $w$ is the wire width, and
estimated the ratio $L/\l_\phi$.
The results of this fitting are represented by the solid curves in Fig.~\ref{fig3}(a).
Here, the fitting range is restricted by the condition that the magnetic length $l_B$ is larger than the width $l_B = \sqrt{\frac{\hbar}{eB}} > w = 100$ nm, a magnetic length range based on the constraint that the magnetic flux through a phase coherent area is less than one flux quantum.
It can be seen from this that the magnetoconductance of a quasi-1D Bi$_2$Se$_3$ wire is accurately explained by  the WAL theory.
The ratio $L/l_\phi $ is plotted as a function of wire length in Fig.~\ref{fig3}(b).
Here, the typical values of  coherence length, as derived from the proportionality coefficient, are $272\pm 8, 243\pm 7$, and $151\pm 4$ nm at $2, 5$, and $12$ K, respectively.
The observed deviation from straight lines demonstrates that $l_\phi$ depends somewhat on the configuration of the voltage probes.
We show the probe configurations and
list the values of $l_\phi$ used for rescaling the $x$-axis of Fig.~\ref{fig4} in Table~\ref{tab:supplementalMatsuo}. The probe configurations are shown in Fig.~\ref{figs1}.
\begin{table}[t]
  \centering 
  \vskip-\lastskip\caption{$L/l_\phi $ at 2 K, 5 K and 12 K.
   The numbers in brackets are values of $l_\phi $ in units of nm.
  The first column specifies the voltage probe configuration, with letters A to E
  corresponding to the marks in Fig. \ref{figs1}.
  The second column $L$ is the distance between the probes.
  The data indicated D-E* was obtained in the different cooling down of the sample.}
\small
\begin{tabular}{|c|c|l|l|l|}
\hline\hline
probe & $L$ ($\mu$m)&$L/l_\phi$ ($l_\phi$) (2K)& $L/l_\phi$($l_\phi$) (5K) &$L/l_\phi$($l_\phi$) (12K) \\\hline
A-B &5.0    &19.2 (260)  &25.9 (192)    &31.1 (160)\\\hline
A-C & 9.0   &31.0 (290)   &35.3 (255)    &50.7 (178)\\\hline
A-D & 12.0 &41.5 (289)   &46.2 (260)    &70.3 (170)\\\hline
A-E & 14.0 &48.1 (291)  &52.4 (267)  &89.4 (157)\\\hline
A-F & 15.0 &52.0 (288)  &56.8 (264)  &102.9 (146)\\\hline
B-C & 4.0   &14.3 (280)   &16.4 (245)    &23.2 (172)\\\hline
B-D & 7.0  &26.3 (266)   &29.0 (241)    &42.4 (165)\\\hline
B-E & 9.0  &35.3 (255)    &40.6 (222)     &61.6 (146)\\\hline
B-F & 10.0 &37.9 (263)   &44.8 (223)     &76.0 (132)\\\hline
C-D & 3.0  &11.8 (255)   &13.5 (222)     &19.0 (158)\\\hline
C-E & 5.0 &24.8 (202)     &26.5 (188)     &36.1 (139)\\\hline
C-F & 6.0 &25.4 (236)     &30.9 (194)     &48.7 (123)\\\hline
D-E & 2.0 &11.6 (173)      &11.9 (168)        &16.6 (120) \\\hline
D-F & 3.0 &17.3 (174)      &17.9 (167)        &29.1 (103)\\\hline
D-E*& 2.0 &11.5 (174)      &11 (180)       &17.0 (118)\\\hline
\end{tabular}
\label{tab:supplementalMatsuo}
\end{table}

\subsection{Quantitative study of the fluctuation amplitude}
\begin{figure}[t]
\begin{center}
\includegraphics[width=.95\linewidth]{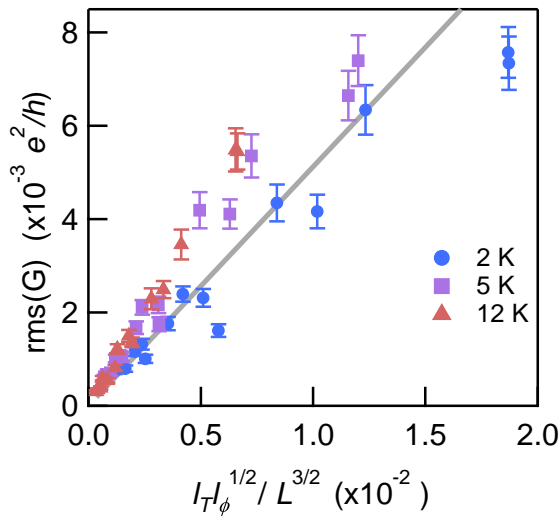}
\end{center}
\vskip-\lastskip \caption{
Rms($G$) as a function of $l_T {l_\phi }^{1/2}/{L}^{3/2}$.
 Values of rms($G$) at 2 (circles), 5 (squares) and 12 K (triangles) increases as $l_T {l_\phi }^{1/2}/{L}^{3/2} $ increases.
The value of rms($G$) is proportional to $l_T {l_\phi }^{1/2}/{L}^{3/2}$, consistent with the predictions of  UCF theory \cite{beenakkerssp1991}.
The straight line represents the relation expressed by Eq.~(\ref{eq1}) with $C=0.51$. }
\label{fig4}
\end{figure}
As a result of the WAL effect, the coherence length will be longer than the thermal diffusion length, which should cause UCF to arise from self averaging and energy averaging; correspondingly,
the UCF amplitude  should satisfy the relation expressed in Eq.~(\ref{eq1}).
The values of rms($G$) at various temperatures as a function of $l_T {l_\phi }^{1/2}/{L}^{3/2}$ are shown in Fig.~\ref{fig4}, where the circles, squares, and triangles represent data obtained at 2, 5, and 12 K, respectively.
We note that, instead of the mean value estimated from the proportionality coefficient of the plot in Fig.~\ref{fig3}(b), the value of $l_{\phi}$ estimated by  fitting to the WAL formula (Fig. \ref{fig3}(a) and Eq.~(\ref{eq2})) is used here (see Table~\ref{tab:supplementalMatsuo}).

As shown in Fig.~\ref{fig4}, the amplitude of  conductance fluctuation is proportional to $l_T {l_\phi }^{1/2}/{L}^{3/2}$, which is consistent with Eq.~(\ref{eq1})
and provides reliable evidence that the origin of the conductance fluctuation is UCF.

We next discuss the value of the coefficient $C$ in Eq.~(\ref{eq1}), for which we estimated a
value of $0.51\pm 0.02$ by the fitting rms($G$) to a linear function of $l_T {l_\phi }^{1/2}/{L}^{3/2}$, as shown in Fig.~\ref{fig4}.
This is consistent with the fact that the fluctuation originates from UCF.
Our result of the coefficient $C$, however, is less than the theoretical value, $\sqrt{\pi/3}=1.023\cdots$ \cite{Akkermans}. 
This discrepancy may be caused by the fact that the condition $l_T\ll l_\phi \ll L$ in the formula Eq.~(\ref{eq1}) is not closely satisfied.  Alternatively, the deviation can be explained in terms of  specific properties of the topological insulator; in either case, further study is needed to clarify the situation.

\subsection{Conclusion}
Together with our previous work \cite{matsuoprb2012}, this study very strongly supports the hypothesis that conductance fluctuation in Bi$_2$Se$_3$ originates from UCF.
Our work has also demonstrated the transport properties of electrons in the bulk conduction band of 3DTI;
this kind of understanding is necessary in determining the transport properties of  disordered Dirac and Weyl electrons \cite{kobayashi13}.

In this study, we investigated the transport properties of  Bi$_2$Se$_3$ wires of various lengths,
determining that the amplitude of conductance fluctuation is proportional to $l_T {l_\phi }^{1/2}/{L}^{3/2}$
with a proportionality constant not far from the value predicted by UCF theory.
This result constitutes reliable evidence that the conductance fluctuation in this material originates from UCF.

This work was partially supported by the JSPS Funding Program for Next Generation World Leading Researchers (GR058), KAKENHI No. 23540376, and Grant-in-Aid for Scientific Research on Innovative Areas "Fluctuation \& Structure" (No.25103003) from the Ministry of Education, Culture, Sports, Science, and Technology of Japan.

\bibliography{ref}

\end{document}